# High-pressure X-ray photon correlation spectroscopy at fourth-generation synchrotron sources

**Antoine Cornet[a]*, Alberto Ronca[a]*, Jie Shen[a], Federico Zontone[b], Yuriy Chushkin[b], Marco Cammarata[b], Gaston Garbarino[b], Michael Sprung[c], Fabian Westermeier[c], Thierry Deschamps[d] and Beatrice Ruta[a]***

[a] Institut Néel, Université Grenoble Alpes and Centre National de la Recherche Scientifique, 25 rue des Martyrs - BP 166, Grenoble, 38042, France

[b] European Synchrotron Radiation Facility, 71 avenue des Martyrs, CS 40220, Grenoble, Rhône-Alpes, 38043, France

[c] DESY, Notkestraße 85, D-22607, Hamburg, Germany

[d] University of Lyon, Université Claude Bernard Lyon 1, CNRS, Institut Lumière Matière, Villeurbanne, F-6922, France

Correspondence email: antoine.cornet@neel.cnrs.fr; alberto.ronca@neel.cnrs.fr; beatrice.ruta@neel.cnrs.fr

**Abstract** A new experimental setup combining X-Ray Photon Correlation Spectroscopy (XPCS) in the hard x-ray regime and a high-pressure sample environment is developed to monitor the pressure dependence of the internal motion of complex systems down to the atomic scale in the multi-gigapascal range, from room temperature to 600K. The high flux of coherent high energy x-rays at 4$^{th}$ generation synchrotron source solves the problems caused by the absorption of the Diamond Anvil Cells used to generate the high pressure, enabling the measurement of the intermediate scattering function over 6 orders of magnitude in time, from $10^{-3}$ s to $10^{3}$ s. The constraints posed by the high-pressure generation such as the preservation of X-ray coherence, as well as the sample, pressure and temperature stability, are discussed, and the feasibility of high-pressure XPCS is demonstrated through results obtained on metallic glasses.

## 1. Introduction

Amorphous materials are ubiquitous in our daily life. Although they lack a well-defined long-range microscopic structure, many hard and soft glassy systems feature common microscopic relaxation processes which control the evolution of their macroscopic properties (Ngai, 2011). Example are proteins in crowded media (Bin *et al.*, 2023; Begam *et al.*, 2020; Roosen-Runge *et al.*, 2011; Foffi *et al.*, 2014), polymers (Conrad *et al.*, 2015; Arbe *et al.*, 1998; Cangialosi, 2014), clays (Angelini *et al.*, 2014; Shalkevich *et al.*, 2007; Jabbari-Farouji *et al.*, 2008; Nigro *et al.*, 2020), viscous alloys (Busch *et al.*, 2007; Wang, 2019; Gallino, 2017), network glasses (Sidebottom, 2015; Micoulaut, 2016) and pharmaceutical compounds (Wang *et al.*, 2021; Rodríguez-Tinoco *et al.*, 2016). Among the large family of disordered systems, structural glasses play a key role, often being considered as archetypes of materials far from thermodynamic equilibrium. A comprehensive microscopic theory of their amorphous state, a long-sought quest in material physics (Gibbs & DiMarzio, 2004; Adam & Gibbs, 2004; Barrat & Hansen, 2003), depends on an accurate description of the system dynamics, i.e. its internal motion, from the inter-constituent's length scale up to the macroscopic regime, and over the complete timescale of the ongoing relaxation processes, i.e. from ps to s (Egami & Ryu, 2020).

This description of the state of the system needs to go through the determination of high-order correlation functions that go beyond the spatially averaged structure factor or pair distribution function. Experimentally, higher order correlation functions and thus dynamical properties of complex systems at nanometric and atomic length scales can be obtained by means of the X-ray Photon Correlation Spectroscopy (XPCS) technique (Sutton, 2008; Madsen *et al.*, 2016; Shpyrko, 2014; Lehmkühler *et al.*, 2021). XPCS quantifies the temporal intensity correlation of fluctuating speckles generated by the scattering from a disordered system, $g_2(q,\Delta t)$, to obtain information on the internal dynamics of the system by the Siegert relation:

$$g_2(q,\Delta t) = 1 + A(q)|F(q,\Delta t)|^2 \quad (1)$$

where $A(q) = \gamma f_q^2(q)$ depends on the non-ergodicity parameter $f_q(q)$ and the degree of coherence $\gamma$ (contrast) of the experimental geometry. $F(q,\Delta t)$ is the intermediate scattering function (ISF), i.e. the Fourier transform of the Van Hove correlation function (Madsen *et al.*, 2016; Shpyrko, 2014; Lehmkühler *et al.*, 2021).

To date, experimental studies of the relaxation dynamics in glass-formers have focused on the temperature dependence of the atomic motion (Amini *et al.*, 2021; Wang *et al.*, 2015), its response to external mechanical and thermal stresses (Zhou *et al.*, 2020; Küchemann *et al.*, 2018; Luo *et al.*, 2020; Das *et al.*, 2020) or to its interaction with intense x-ray beams in the case of oxide and chalcogenide glasses (Martinelli *et al.*, 2020; Alfinelli *et al.*, 2023; Dallari *et al.*, 2023; Chushkin, 2020; Li *et al.*, 2022; Pintori *et al.*, 2019).

Among the different properties of interest affecting the dynamics of glasses and liquids, density plays a major role. The viscosity of glass formers strongly depends on the density in molecular liquids and polymers (Grocholski & Jeanloz, 2005; Kondrin *et al.*, 2012; Paluch *et al.*, 2007), which in turn affects the glass transition (Paluch *et al.*, 2001; Niss & Alba-Simionesco, 2006; Niss *et al.*, 2007). Density also affects relaxation phenomena deep in the glassy state: physical aging, that is the slow relaxation of the glass towards a metastable equilibrium state, was shown to be mediated by density-driven rearrangements releasing residual stresses and medium range ordering processes not affecting the local density in metallic glasses (Giordano & Ruta, 2016).

Transitions between different amorphous states have been also reported in many out-of-equilibrium materials, where pressure can drive the system from a low to a high density amorphous state with different physical properties (Tanaka, 2020; Machon *et al.*, 2014; Zhang *et al.*, 2010). These liquid-liquid or glass-glass polyamorphic transitions appear in all kind of systems, including the canonical case of water (Mishima, 2021;

Amann-Winkel *et al.*, 2013), covalent (Machon *et al.*, 2014), ionic (Wojnarowska *et al.*, 2022) and metallic systems (Sheng *et al.*, 2007). While structural studies under pressure have been reported so far, very little is known on the evolution of the relaxation dynamics during pressure-induced polyamorphic transitions due to the experimental challenge behind the use of high pressure sample environments for dynamical studies, and previous studies have focused mainly on temperature induced liquid-liquid transitions (Perakis *et al.*, 2017; Amann-Winkel *et al.*, 2013; Hechler *et al.*, 2018).

In the case of XPCS, the relatively low coherent flux at photon energies, $E$, higher than $E$ >15 keV in the majority of 3rd generation synchrotrons limited the use of bulky sample environments as those necessary for studies under high pressure (HP). Thanks to the advent of 4th generation synchrotrons, such as the extremely brilliant source ESRF (ESRF-EBS), it is possible to deliver high coherent x-ray fluxes also at high photon energies (E~20 keV), solving the absorption issue related to bulky sample environments and thereby opening the way to time-resolved, high quality HP-XPCS in glassy systems, unlocking a new field of investigations (Cornet *et al.*, 2023; Zhang *et al.*, 2023). Based on the first HP-XPCS measurements, this paper addresses the different requirements and main issues encountered at the experimental level to obtained reliable dynamical data under high pressure and high temperature with atomic scale XPCS.

## 2. Experimental

### 2.1. Signal-to-noise ratio, absorption and coherence at 4th generation synchrotron sources

The signal to noise ratio (SNR) in an XPCS experiment scales linearly with the average intensity $\langle I(q)\rangle$ and the square root of the minimum sampling time $\tau_{min}$ and measurement time $t$ and is being defined as:

$$SNR = \langle I(q)\rangle \frac{A(q)}{\sqrt{1+A(q)}}\sqrt{\tau_{min} \times t \times N_{px}} \tag{2}$$

where $N_{px}$ is the number of pixels of the detector used to average the correlation function (Jankowski *et al.*, 2023). Relation (2) has three remarkable consequences. SNR will depend linearly with the coherence flux (and not with its square root, as for standard intensity measurements) and any gain factor $n$ will readily translate into a possible minimum sampling time $\tau_{min}$ smaller by a factor $n^2$ for the same SNR. On the other hand, decreasing the intensity by a factor $\alpha$ translates into an increase of the total acquisition time $t$ by $\alpha^2$ for a similar SNR. An additional flux reduction by an absorbing sample environment implies even longer acquisition times, rapidly exceeding laboratory time scales.

For synchrotron radiation sources the coherent flux, $F_c$, relates to the source brilliance, $B$, as $F_c = B(\lambda/2)^2$, where $B$ quantifies the emission of photons per unit time, unit area, unit solid angle and band-width. The $\lambda^2$ dependence causes a strong decrease of coherent flux at high energies, which can be compensated by a large brilliance. Practically, this means that at 3rd generation synchrotron radiation (SR) sources XPCS was usually applied with a coherent flux in the order of $5{\cdot}10^9$-$5{\cdot}10^{10}$ photon/s in the 8-11 keV energy range (beamlines ID10 at ESRF (Favre-Nicolin *et al.*, 2017), 8-ID-E at APS (Zhang Jiang *et al.*, 2023), P10 at PETRA III (Michael Sprung *et al.*, 2023)). This is an energy domain where the absorption of even a few millimetres of structural material is prohibitive to obtain a satisfying signal to noise ratio over a reasonable time scale. Coherent x-rays are still available at higher energies (Frost *et al.*, 2023), albeit at such substantial reduced fluxes, that in some cases the use of the bulky sample environment used for high-pressure generation becomes non-feasible.

The advent of the 4th generation SR sources like the ESRF-EBS led to a considerable jump in brilliance, and therefore in coherent flux (Favre-Nicolin *et al.*, 2017), by up to two orders of magnitude thanks to the reduction of the horizontal emittance of the electron beam (Raimondi *et al.*, 2023). As example, the table I lists the brilliance from the X-ray beam produced by the U27 undulator at the ID10 beamline at EBS-ESRF, computed from the experimentally measured spectral flux[1] (Zontone *et al.*, 2010) over the first and third harmonics. Pre-EBS, high-β electron source reference values are reported for comparison. A 70x gain is observed at the third harmonic at 21 keV (Jankowski *et al.*, 2023), resulting in an available coherent flux of about $10^{11}$ ph/s, exceeding even the maximum coherent flux previously available from the full ID10 high-β straight section (consisting of two U27 undulators and one U35 undulator) at 8 keV (Jankowski *et al.*, 2023).

On the same table, we report the transmission through a typical Diamond Anvil Cell (DAC), the apparatus often used to generate high pressure. This transmission, calculated through two 1.7 mm thick diamonds, increases from 0.5% to 68% from 8 keV to 21 keV, and shows how the energy shift opens the door to high-pressure measurements based on coherent scattering.

In addition to the increased incident coherent flux, several other constraints exist to perform HP-XPCS, including the preservation of the coherence through the sample environment, the pressure and temperature stability, the absence of shear stress on the sample at high pressure and the impact of the intense irradiation on the sample environment. These will be discussed later.

| | **old ESRF** | | **ESRF-EBS** | | **Transmission** |
|---|---|---|---|---|---|
| | E (keV) | Brilliance (ph/s/mm$^2$/mr$^2$/0.1%bw) | E (keV) | Brilliance (ph/s/mm$^2$/mr$^2$/0.1%bw) | |
| **1st harmonic** | 7.93 | $2.43\times10^{19}$ | 7.22 | $1.26\times10^{21}$ | 0.5% |
| **3rd harmonic** | 24 | $8.25\times10^{18}$ | 21.88 | $6.18\times10^{20}$ | 68% |

**Table 1** Peak energy and brilliance of the 1st and 3rd harmonic of a single U27 (1.5 m long) undulator at the ESRF ID10 beamlime, before and after the ESRF-EBS upgrade, with the transmission of the x-ray through a typical diamond anvil cell at the corresponding energy.

### 2.2. Setup and feasibility

The experimental setup used to probe the effect of pressure on the internal atomic dynamics of disordered systems is schematized in the Fig. 2a. The incoming beam is produced by three undulators, followed by a first collimation with high power slits and a Pd-coated double mirror at grazing incidence to supress higher harmonics. The x-ray beam is made monochromatic by a cryo-cooled channel cut Si(111) monochromator ($\Delta E/E = 1.4\times10^{-4}$) and can be focussed by three independent sets of Be Compound Refractive Lenses (CRLs), located at 36m, 52.2m and 56.3m from the source (Jankowski *et al.*, 2023). The third focus stage, shown in Fig. 2, consists of an array of nine Be CRLs of radius 100µm, and leads to a beam size of 5.4x1.6 $\mu m^2$ (HxV, FWHM) at the sample position 61m downstream the source.

[1] With primary slits (at 27.2 from the source) open to $0.15\times0.15$ mm$^2$.

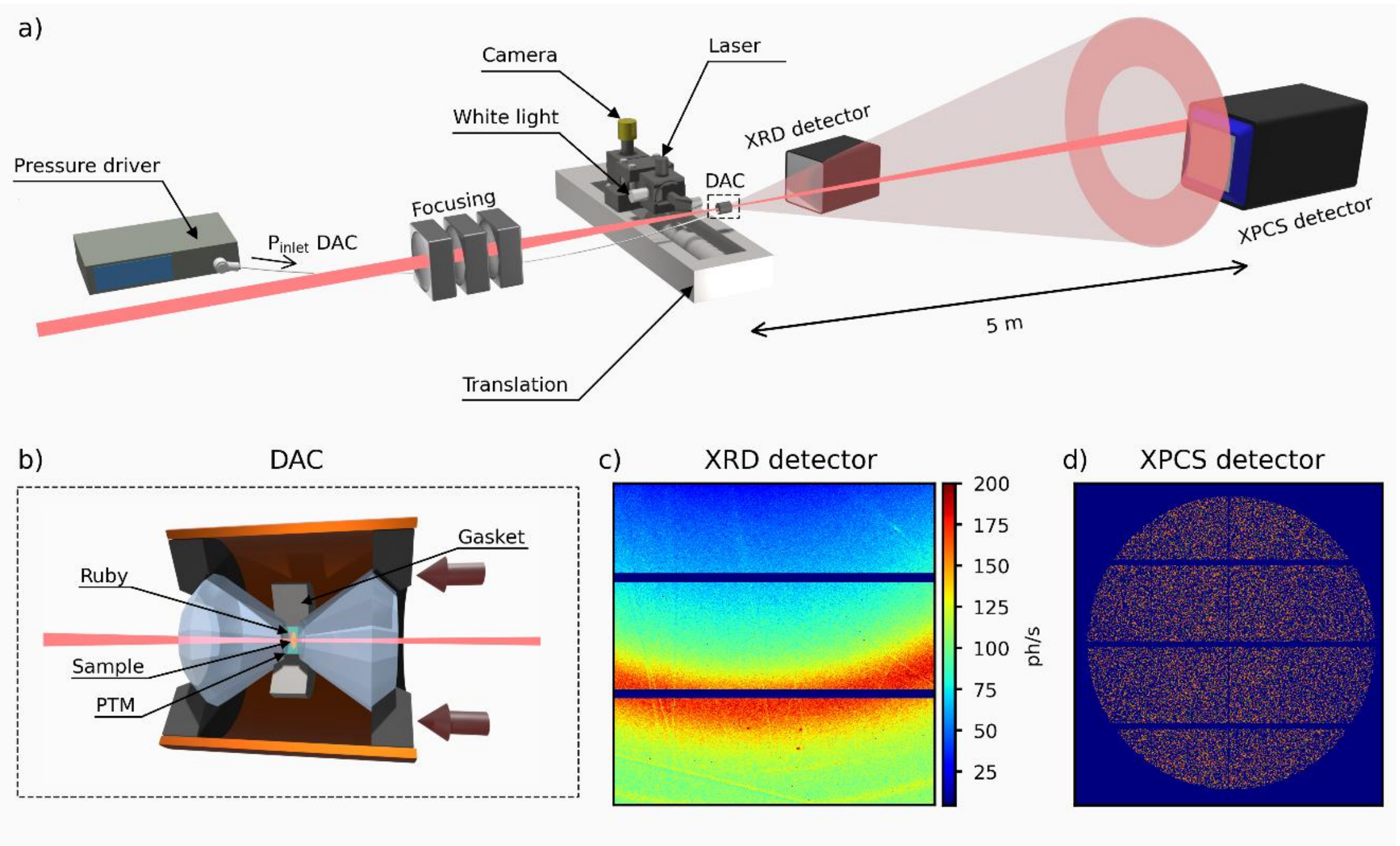

**Figure 1** Schematic of the wide-angle High-Pressure High-Temperature XPCS experimental setup (a), with a sketch of the Diamond Anvil Cell (b), and raw images from a PILATUS 300K detector dedicated to diffraction (c) and an EIGER 4M to XPCS (d). The integration time for the diffraction and XPCS images are 60s and 0.1s respectively. A simplified color code is used for the XPCS detector, where all active (non-zero) pixels are shown, corresponding to 1 (96.67%), 2 (3.25%) or 3 (0.08%) impinging photons. All the elements necessary for HP(-HT) XPCS are labelled.

The large sample-to-detector distance is constrained by the speckle size $d_{speckle} \approx \lambda \times R/s$, $R$ and $s$ being the sample-to-detector distance and the X-ray spot size respectively, which must match the detector pixel size to resolve the speckles and collect the highest scattered intensity for the optimum SNR [23]. For these experiments, data are collected by a CdTe EIGER 4M detector from DECTRIS, with a pixel size of 75×75 $\mu m^2$. The correlation functions are calculated from time series of the scattering patterns (Fig. 2d) using the event correlator described in (Chushkin *et al.*, 2012). The detector covers a limited range in scattering vector q and the correlation functions are calculated and averaged over azimuthal sectors in q where the scattering intensity is considered constant. In order to characterize the structure of the amorphous material simultaneously to the dynamics, an additional PILATUS 300k detector is placed downstream from the sample for standard X-ray diffraction. The diffraction pattern, shown in the Fig. 2c, spans a sufficiently wide range in q to cover the first two diffraction peaks, which allows to follow and analyse the evolution of the structure factor during the XPCS measurements, as shown in Fig. 3a.

To generate the high-pressure, a membrane-based Diamond Anvil Cell is used (Fig. 2b). Two diamonds facing each other form an experimental volume of dimensions 70×300 µm (height×diameter), contained radially with a metallic gasket previously deformed at the target pressure. In this DAC the pressure is generated by a metallic membrane. One diamond is attached to a mobile piston, driven by the metallic membrane which inflates when pressurized. The relative change in diameter from the membrane (50 mm wide) to the culet size of the diamond (600 µm wide) generates pressures in the multi-GPa range for input pressures within 0 to 100 bars. The membrane pressure is controlled remotely by an automatic pressure driver (PACE 5000, Druck) with a precision of $10^{-3}$ bars. A Pressure Transmitting Medium (PTM) filling the experimental chamber ensures a hydrostatic pressure on the sample, although deviatoric shear stress inevitably appears above the solidification of the PTM (2 to 20 GPa at 300K depending on the PTM (Klotz, Chervin *et al.*, 2009)). In this study, we used a *Le Toullec* type membrane driven DAC equipped with 3.1 mm wide, 1.7 mm high, 600 µm culet size diamonds, and samples of about 50 µm in size.

The experimental chamber also comprises of a fiducial marker for the pressure determination. We used ruby spheres with a controlled Cr amount, as the evolution of the $^2E \rightarrow ^4A_2$ transition wavelength is well calibrated under pressure (Shen *et al.*, 2020). Pressure determination is also possible from the cell parameter of pure compounds with known equation of state, provided that enough reflections are visible within the XRD detector across the full pressure range. An optical bench which translates in and out of the beam path provides the 405 nm laser excitation and collection for the ruby signal through a 10× magnification objective, dispersed by a 600 lines/mm grating and recorded by a Peltier cooled-ccd camera for a final resolution of 0.1 nm. This optical bench also provides a white light illumination of the DAC volume collected by a camera for an online diagnostic of the sample environment alternatively to the x-rays.

The effect of the DAC, or more specifically the diamonds, on the speckle visibility is shown in the Fig. 2. In this figure, we plot the data obtained on a $Pt_{42.5}Cu_{27}Ni_{9.5}P_{21}$ metallic glass, compressed in-situ at 1.5 GPa, using 4:1 Methanol:Ethanol mixture as PTM, which ensures hydrostaticity at the applied pressure. The total scattered intensity during the scan (7000 frames, 0.1s acquisition time) in the detector centred on q = 2.73 Å$^{-1}$ is visible in Fig. 2b,c. It features the top of the broad first diffraction peak of the glass, plus straight lines which correspond to the Kossel lines from the monocrystalline diamonds (Faigel *et al.*, 2016; Gog *et al.*, 1995).

The dynamics of the glass is represented by the $g_2(q,\Delta t)$ function obtained by the correlation of the 7000 frames in the panel d. One curve is obtained by considering the full active area of the detector (all the active pixels in the panel c), the other with the pixels corresponding to the Kossel lines excluded from the correlation calculation. A similar well-defined $g_2(q,\Delta t)$ function is obtained with a high SNR in both cases, only shifted vertically. This indicates that the Kossel lines only add a static contribution to the correlation function, and do not affect the probed dynamics of the glass. However, in the case of slow dynamics, full decorrelation is not always resolved in the duration of the measurement. If the reference value for complete decorrelation is affected by static correlations such as the Kossel lines (or other spurious scattering), the corresponding constrained parameter in the modelling of the data can be erroneous. In this case, the static reference signal acquired next to the sample can be used to constrain the modelling of the data with the correct baseline.

The dashed lines correspond to data taken while the x-ray beam points on the diamond only, outside the sample. Contributions to this flat correlation function include x-rays scattered by the PTM and the two diamonds. Since the used PTM is a liquid (alcohol), the timescale of the ISF of the alcohol mixture is many orders of magnitude faster than our temporal resolution of 0.1s, so only the incoherently scattered x-rays contribute to the dashed flat signal of Fig. 2d. Scattering from diamonds is purely static, so the dashed lines represent the baselines of the previous $g_2(q,\Delta t)$ functions, and allow for a complete parametrization of the data modelling. To compare the effect of the DAC on this reference correlation value, we plot in the same panel the $g_2(q,\Delta t)$ function obtained in a furnace (i.e. no DAC) for a $Pt_{42.5}Cu_{27}Ni_{9.5}P_{21}$ metallic glass sample from the same batch, at a temperature where the timescale of the $g_2(q,\Delta t)$ function is similar. Once the mask is applied, both $g_2(q,\Delta t)$ functions have a comparable baseline value after full decorrelation, which demonstrates that the contribution from the DAC to the correlation function is not significant when crystalline features of the scattering signal are properly masked.

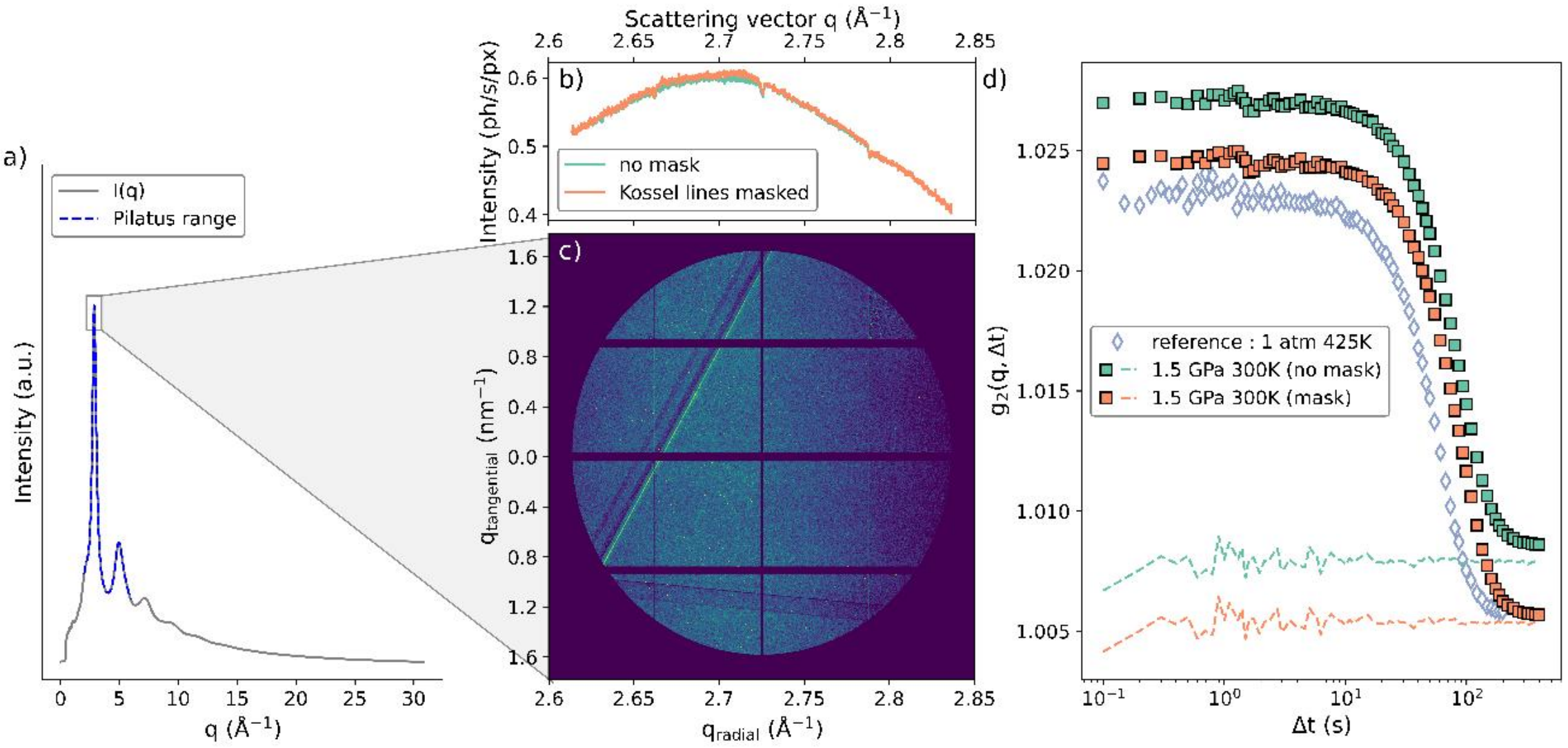

**Figure 2** a) Diffracted intensity of a $Pt_{42.5}Cu_{27}Ni_{9.5}P_{21}$ metallic glass, taken at 300K and 1 atm at the ID15a beamline of the ESRF. The range covered by the diffraction detector is shown by the blue dashed line. The range covered by the XPCS detector is indicated by the box on the first peak. b) Scattered intensity after azimuthal integration of the scan-averaged intensity within the XPCS detector, considering all active pixels in the detector (orange solid line), or after masking the two large Kossel lines (green solid line). c) The averaged two-dimensional pattern measured by the XPCS detector during a 700s scan. d) $g_2(q,\Delta t)$ functions obtained from the same scan, with and without masking the Kossel lines (full symbols), and the $g_2(q,\Delta t)$ function generated by the diamonds and PTM, obtained when aiming the x-ray beam out of the sample (dashed lines). The $g_2(q,\Delta t)$ function of a similar metallic glass obtained in a furnace is shown for comparison (empty symbols).

The similar amplitude, also called contrast, of the $g_2(q,\Delta t)$ functions obtained in-situ at 1.5 GPa and at 1 atm demonstrates also the absence of any degradation of the coherence of the x-ray beam by the diamonds. The slightly higher contrast at 1.5 GPa originates either from geometrical effects as the sample gets thinner with pressure, or to pressure-induced structural changes in the material through the non-ergodicity parameter $f_q$. Overall, the diamonds and the pressure transmitting medium do not have a significant impact on the contrast of the correlation functions. The relaxation phenomena behind the atomic dynamics, and therefore behind the long-time decay of the ISF in Fig 2d are described in details in (Cornet *et al.*, 2023). There, it was shown that a compression at moderate pressures leads to fast intermittent dynamics, and to physical aging at higher pressure. This two-step scenario where the nature of the dynamics changes as a function of pressure supports a rejuvenation and strain hardening observed macroscopically (Pan *et al.*, 2018, 2020) and are consistent with numerical simulations showing the existence of a pressure-induced second local minimum after the first coordination shell which disappears at higher pressure (Ngan *et al.*, 2021).

### 2.3. 3rd and 4th generation synchrotron sources

While the new coherent properties of 4th generation synchrotron sources are ideal for HPHT-XPCS studies, it is important to point out that HP-XPCS can still be performed in some 3rd generation synchrotron sources, such as PETRA III, where XPCS can be performed up to 15 keV. This is highlighted in Fig. 3, where we compare $g_2(q,\Delta t)$ functions obtained under pressure at room temperature on two different metallic glasses at the P10 beamline at the PETRA III synchrotron source (Hamburg, Germany) with data obtained at the ID10 beamline at the ESRF at similar pressures, for samples with identical thicknesses. The SNR in the $g_2(q,\Delta t)$ function at 15 keV remains satisfactory, especially for the $Pt_{42.5}Cu_{27}Ni_{9.5}P_{21}$ glass, demonstrating also the feasibility of HP-XPCS at 3rd generation sources. However, a drop of contrast is visible in the results obtained at 15 keV compared to that obtained at 21 keV. The contrast drop is even more pronounced when considering the respective beam sizes: 2.3x1.4 $\mu m^2$ at 15 keV against 5.4x1.6 $\mu m^2$ at 21 keV (HxV, FWHM), which should translate to a higher speckle contrast at P10 for an identical configuration. As the photon energy effect on the overall contrast depends on many factors, such as a change in the longitudinal coherence length, the speckle size, or a lower scattering angle for the same q value the comparison of data acquired at different energies is not straightforward. The energy shift also leads to a different relative strength in the scattering of the sample and the diamonds, which also affects the contrast. In addition to this multifactorial variation of the contrast with the energy, one must also consider the differences of the experimental setup. In particular, the P10 beamline is equipped with a Si based detector, while the scattering patterns at ID10 are recorded with a CdTe based detector. The quantum efficiency of the former decreases from 98% at 8 keV to 47% at 17.5 keV, while that of the CdTe based sensor remains above 90% below 26 keV, effectively leading to a higher contrast for CdTe based detectors in high energy XPCS. Finally, these detectors can discriminate between photons having energies below or above a defined threshold, but this threshold should not exceed 80% of the used photon energy. Given the multiple edges in the atomic form factors of Pt and Au between 13 keV and 14.5 keV, the loss of contrast is also explained by the contribution of the fluorescence photons that cannot be filtered out in Si based detectors when working at 15 keV, but can be excluded in CdTe detectors working at 21.7 keV.

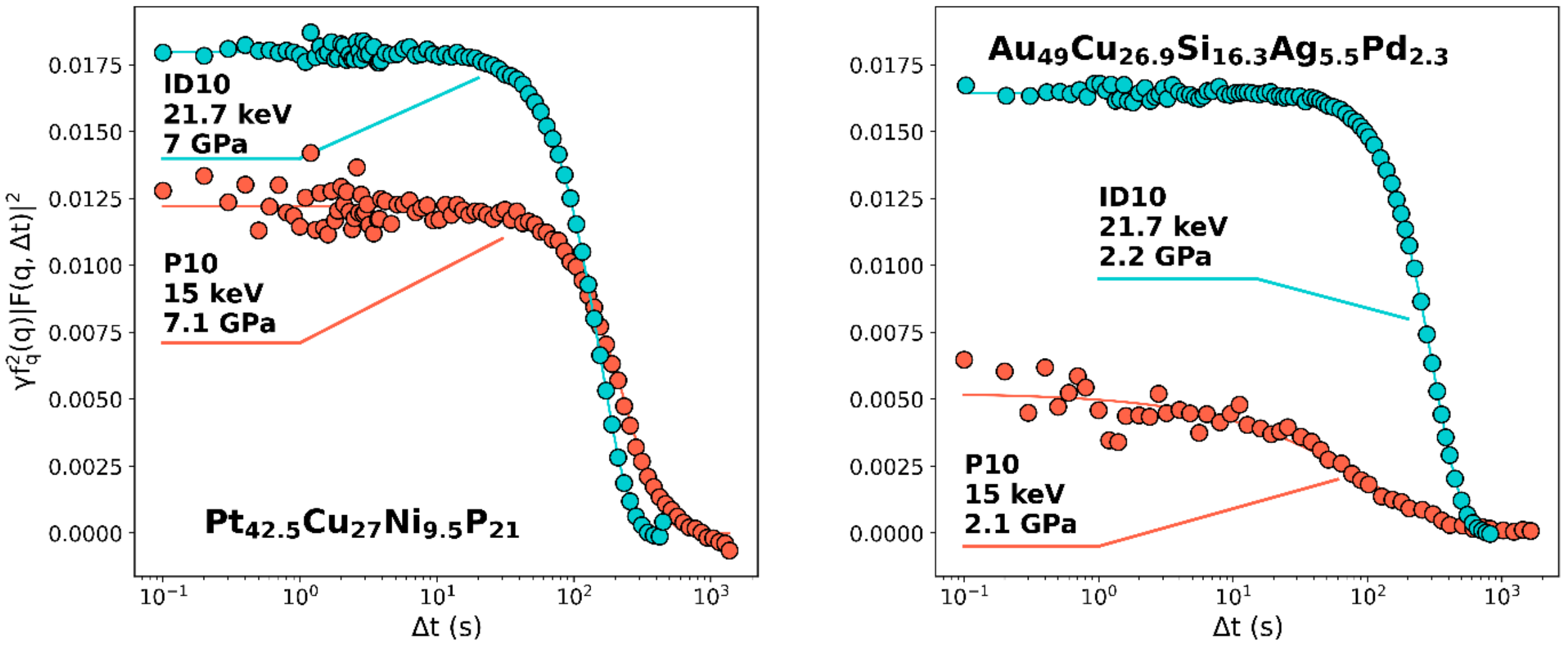

**Figure 3** Intensity-intensity correlation function measured on $Pt_{42.5}Cu_{27}Ni_{9.5}P_{21}$ and $Au_{49}Cu_{26.9}Si_{16.3}Ag_{5.5}Pd_{2.3}$ ribbons at 7 and 2 GPa respectively at the P10 beamline of the PETRA III source, and at the ID10 beamline of the ESRF. For both samples, the ribbons were prepared from the same batch, and thinned to the same thickness (around 15 microns).

This loss of contrast can become a limiting factor. This is particularly true for HPXPCS studies in the supercooled liquid state, where the contrast diminishes even further (Amini *et al.*, 2021).

### 2.4. Pressure and temperature stability

Any movement of the sample position can induce an artificial decorrelation of the successive collected diffracted patterns, as the illuminated volume changes and a new configuration of scatterers is probed. If the timescale of the sample drift is faster or comparable to that of the atomic motion, the probed dynamics correspond to this sample displacement (Busch *et al.*, 2008). Stability is therefore critical as it can be difficult to disentangle artificial from intrinsic dynamics, as for instance in the case of a continuous drift. The horizontal or vertical beam size, $L_{h,v}$, in XPCS measurement is typically around few microns, and glassy dynamics have relaxation times $\tau$ spanning from hundreds of seconds to hours at room temperature and atmospheric pressure (Ruta *et al.*, 2012; Giordano & Ruta, 2016), to the sub-second timescale in the supercooled liquid state (Amini *et al.*, 2021). Therefore, XPCS requires a sample stability better than ~10% × $L_{h,v}/\tau$, which can be difficult to obtain in a DAC, where a slow pressure stabilization usually takes place.

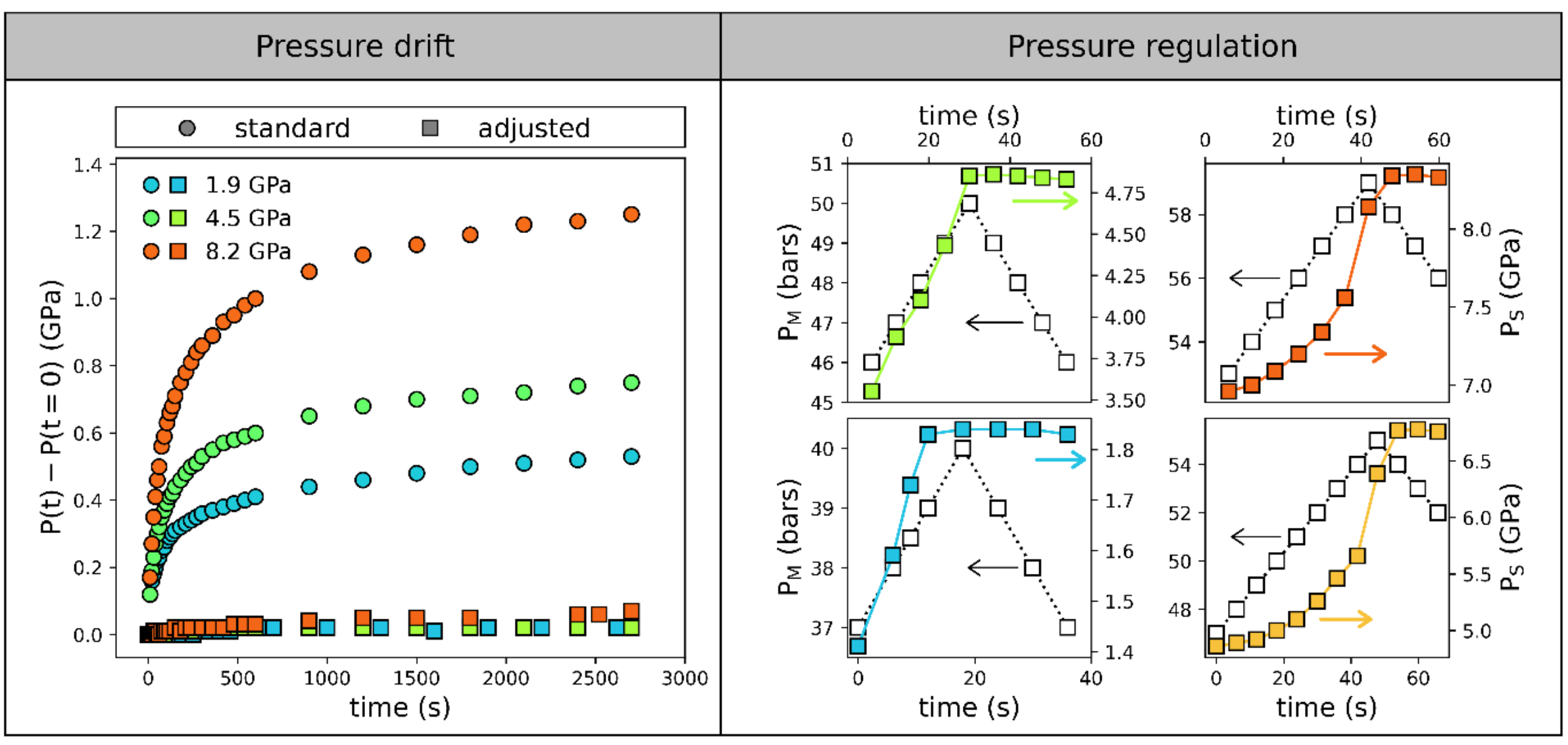

**Figure 4** a) Evolution of the pressure measured on a ruby sphere after stabilization of the membrane pressure for different nominal pressures, with (squares) or without (circles) adjusted protocol. b) Evolution of the sample pressure ($P_S$) after decreasing the membrane pressure ($P_M$) by 4 to 5 bars, at the nominal pressures of 1.8, 4.8, 6.7 and 8.3 GPa.

This is evidenced in the Fig. 4, where we report the pressure stabilization in a DAC after reaching a setpoint on the membrane pressure. All data was obtained in a DAC equipped with 600 µm diamonds (culet size), stainless steel gaskets, 4:1 Methanol:Ethanol mixture as PTM, and a similar rate of 0.1 bar/s on the membrane. The pressure measured within the cell at the moment the membrane pressure is fixed and is given in the legend, and we measure the subsequent evolution of pressure inside the DAC as a function of time. Without regulation (Fig. 4a, circles), both the time needed for equilibration and the amplitude of the pressure drift increase with the nominal pressure, with a drift larger than 1.3 GPa visible for a nominal pressure of 8.2 GPa. More importantly, equilibrium is still not reached in an hour at this pressure, which precludes any XPCS measurement in this timescale. A possibility to improve the timescale and intensity of the pressure equilibration is to decrease slightly the pressure in the membrane after reaching the setpoint, as illustrated in the second panel of Fig. 4. The pressure within the cell is measured while the pressure on the membrane reaches the setpoint and decreases immediately after (empty symbols). It appears that the sample pressure decreases by a maximum of 0.03 GPa for a reduction of the membrane pressure by 4 bars at all pressures. The long-term evolution of the pressure with this dedicated protocol is shown together with the initial pressure drift in the Fig 4a (squares): an obvious reduction of the pressure variation during the equilibration is visible, making XPCS measurement possible almost as soon as the pressure is reached. The evolution of the atomic dynamics under pressure in a $Pt_{42.5}Cu_{27}Ni_{9.5}P_{21}$ metallic glass shows how this can be critical as, at low pressure (<1 GPa), a variation of pressure of only 0.1 GPa leads to an acceleration of the dynamics by a factor 2 (Cornet *et al.*, 2023). Therefore, only a dedicated pressure protocol allows time resolved XPCS measurement under pressure in the 0-10 GPa range tested here, and to pressures up to 30 GPa with carefully selected PTMs (where the pressure standard deviation $\sigma_P$ of 0.1 GPa in He is still reasonable for high Bulk modulus materials).

As for standard x-ray diffraction or spectroscopies techniques, it is also possible to change the temperature along with pressure, opening the possibility to perform high pressure – high temperature XPCS (HPHT-XPCS). Usual means to control the temperature of the samples inside a DAC are laser-heating (Anzellini & Boccato, 2020), internal (Heinen *et al.*, 2021; Mijiti *et al.*, 2020) and external resistive heating (Santoro *et al.*, 2020), the latter being the most limited in terms of temperature range and heating rates, but also the most stable in terms of thermal fluctuations and consequently the best choice for HPHT-XPCS.

As for pressure, a high thermal stability of the whole system is needed to extract reliable dynamics: the stability of the sample temperature is a necessary but not sufficient criterion for HPHT-XPCS, as an excellent temperature regulation could anyway be accompanied by variation in the sample position at the micrometre scales which would partially (or globally) decorrelate the XPCS signal. The mechanical stability at the micrometre level is thus a requirement of XPCS studies which is often not necessary for other techniques, like for instance HP-XRD experiments on glasses. To connect the thermal fluctuations to its effects on the probed dynamics, we report in the Fig. 5 the evolution of the temperatures measured on the heater resistance and at the sample position, and the instantaneous intensity-intensity correlation map $C(t_1,t_2)$ defined by

$$C(q,t_1,t_2) = \frac{\langle I(q,t_1)I(q,t_2)\rangle}{\langle I(q,t_1)\rangle\langle I(q,t_2)\rangle} \quad (3)$$

which represents the correlation between the intensity $I$ measured at a given wave vector $\boldsymbol{q}$ of speckle patterns collected at two distinct times $t_1$ and $t_2$, and the average is performed over all pixels of the detector corresponding to the same $q = |\boldsymbol{q}|$. Averaging correlation values $C_I$ over all identical delay times $\Delta t = t_2 - t_1$, one gets the intensity-intensity correlation function $g_2(q,\Delta t) = \langle C_I(q,t,\Delta t)\rangle_t$.

Focusing on the TTCF on the right panel of the Fig. 5, the main diagonal corresponds to $t_1=t_2=t$, and represents the reference time of the laboratory. The continuous decrease from high (red) to low (blue) correlation values out from this main diagonal corresponds to the $g_2(q,\Delta t)$ function defined in eq. 1, for increasing reference times $\Delta t=t_2-t_1$, the width of the red contour is proportional to the characteristic relaxation time $\tau$.

On the left panel on the Fig. 5, we report the temperature stability at 5 GPa and 539K for a DAC equipped with an external resistive heating collar, where the temperature regulation is performed on the sample temperature (measured by a thermocouple in contact with the back side of the diamond) in a pulsed mode (0 or 100% of heating power), i.e. with a standard commercial set-up usually employed in HP and HT studies.

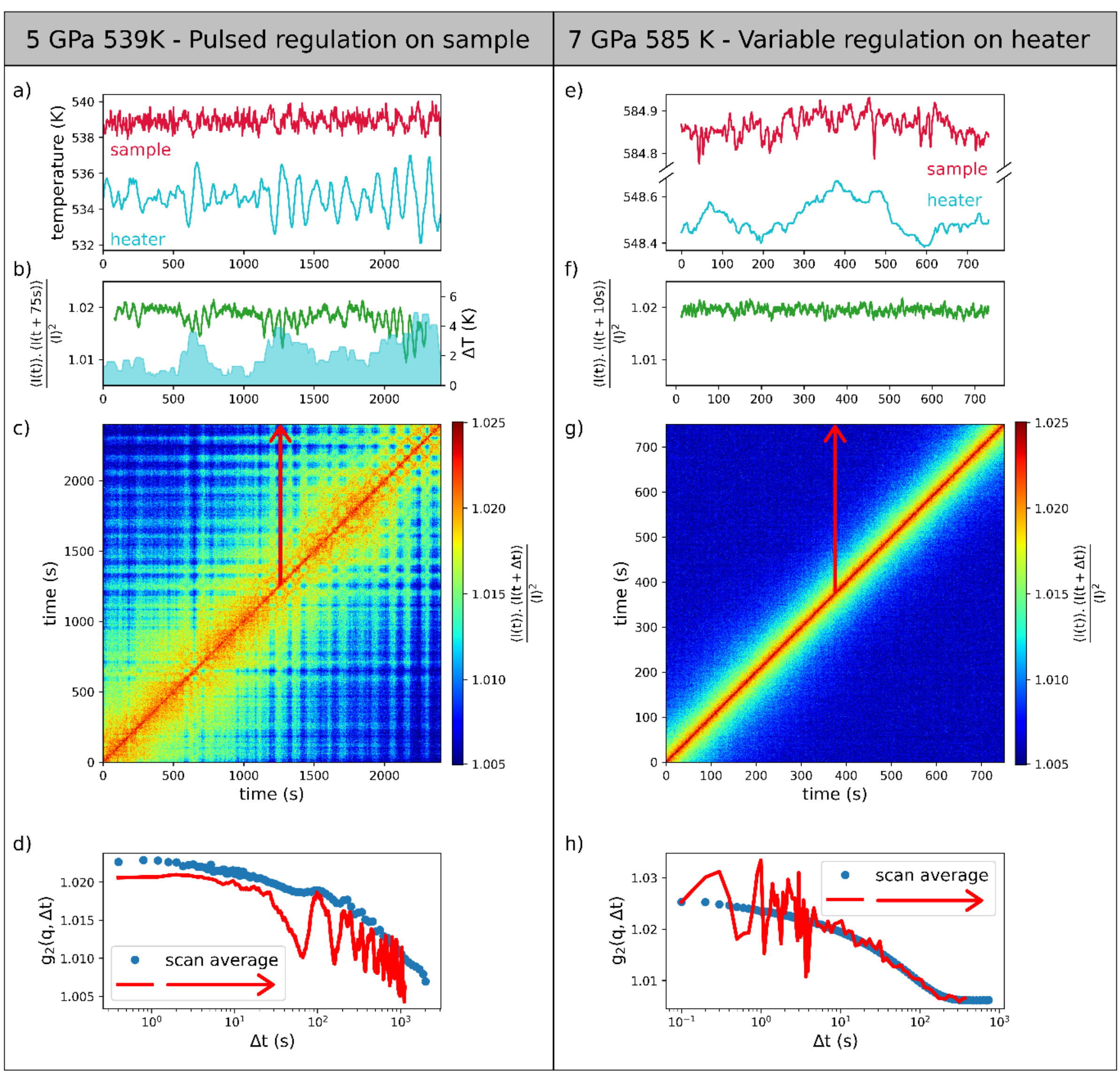

**Figure 5** Effect of the temperature regulation on the sample stability. Left: temperature regulation on sample temperature, pulsed power on heater. Right: temperature regulation on heater temperature, variable power on heater. a,e) furnace and sample temperature, measured by two distinct thermocouples b,f) Correlation values with fixed delay time Δt of 75s and 10s (green), and amplitude of the variation of the furnace temperature. c,g) TTCFs acquired during the temperature monitoring of a,e. d,h) $g_2(q,\Delta t)$ correlation functions (red line) from a fixed reference frame (see the red arrows on the TTCF) compared to those (blue dots) averaged over the measuring time interval.

The sample temperature is stable up to 1K, while the temperature on the heater fluctuates with a peak-to-peak amplitude reaching 5K. The TTCF recorded simultaneously on a $Pt_{42.5}Cu_{27}Ni_{9.5}P_{21}$ metallic glass is also shown on the same temporal axis, and clearly shows a hashed texture for the correlation values. This effect arises from micrometre movements of the sample following the abrupt thermal expansions of the heater due to the pulsed regulation. This movement of the sample position is strictly verified when we compare the amplitude of the oscillation of the furnace temperature ΔT (cyan curve) with the correlation values obtained with a fixed delay time of 75s, where the changes in the correlation values is maximal (green curve): large oscillations in $\langle I(t)\rangle\langle I(t+75s)\rangle/\langle I\rangle^2$ and maxima in ΔT are simultaneous.

In the Fig. 5f, we report the temperature stability and TTCF obtained with the same resistive heating sleeve setup on a similar $Pt_{42.5}Cu_{27}Ni_{9.5}P_{21}$ glass at 7 GPa and 585K, with a dedicated power supply and regulation performed on the heater element with an adjustable heating power after an optimization of the PIDs (Proportional-Integral-Derivative). Not only the temperature stability is enhanced to a peak-to-peak amplitude below 0.1K, but the sample stability is now achieved at the micron scale. The corresponding smooth TTCF demonstrates the feasibility of HTHP-XPCS with an intensity stability characterized by a standard deviation of $7x10^{-4}$ (Fig. 5f). It is important to note that the effect of the sample instability is better highlighted in the TTCF as it is smeared out in the averaged $g_2(q,\Delta t)$ as shown in the bottom panels of the Fig. 5, where we compare the $g_2(q,\Delta t)$ averaged over the full time series to the $g_2(q,\Delta t)$ corresponding to the temporal evolution of the intensity-intensity correlation value from a fixed reference frame (see the red arrows on the TTCF). In the first scenario where the temperature regulation is not optimized, the oscillations appear clearly on the single $g_2(\Delta t)$, but smear out on the averaged correlation function. In the case where the temperature regulation is optimized, dynamics is homogeneous over the duration of the scan, and the two $g_2(q,\Delta t)$ functions coincide as in Fig. 5h. Therefore, the consistency of the results should always be verified from the TTCF with a high intensity resolution as lower x-ray fluxes can hide specific features in the TTCFs.

Importantly, the pressure on the sample changes with temperature, and a membrane driven DAC with a remote-controlled pressure inlet combined with continuous monitoring of the sample pressure is necessary to maintain a pressure stability upon heating or cooling, through a manual compensation of the temperature induced variation of the sample pressure.

Finally, temperature also promotes the creep deformation of the gasket under pressure, which potentially translates to sample movements if the sample is in contact with the gasket, and to a pressure drift within the measurement. The first issue disappears when the sample is positioned in the centre of the experimental volume, with no contact with the gasket. Regarding the pressure drift, the protocol described in the Fig. 4 leads to pressure uncertainties lower than 0.2 GPa, even at temperatures as high as 630K.

### 2.5. Pressure transmitting medium (PTM)

The choice of the pressure transmitting medium depends on the pressure-temperature path taken during the measurement, as the stress exerted on the sample must remain hydrostatic to avoid plastic flow within the probed material. Depending on the pressure and temperature range investigated, a specific PTM is chosen considering its degree of hydrostaticity (Klotz, Chervin *et al.*, 2009; Klotz, Paumier *et al.*, 2009) or its phase diagram (Young *et al.*, 1987; Datchi *et al.*, 2000; Vos *et al.*, 1991). In addition to the strict hydrostaticity constraint, molecular PTMs can also be affected by the strong x-ray beam of the new generation of synchrotron sources. In particular, the 4:1 Methanol:Ethanol mixture shows degassing under irradiation at low pressure, as shown in the Fig. 6. The micrograph of the loaded cell at P = 0.45 GPa shows the position of the sample and of several rubies within the 300 μm hole of the stainless-steel gasket (label 1). After successive scans corresponding to a dose of $5.4\times10^9$ Gy, we observe the apparition of three bubbles, clearly visible in the new micrograph (label 2), which are resorbed when increasing the pressure to 1.35 GPa (labels 3 and 4). The same pattern appears at this new pressure step for an even lower dose: numerous small bubbles are visible in the micrograph taken after an additional deposited dose of $6.5\times10^8$ Gy (label 5), which are again resorbed when increasing pressure to 2 GPa (label 6). No bubbles ever appeared above 1.4 GPa, even for doses up to one order of magnitude above these considered here.

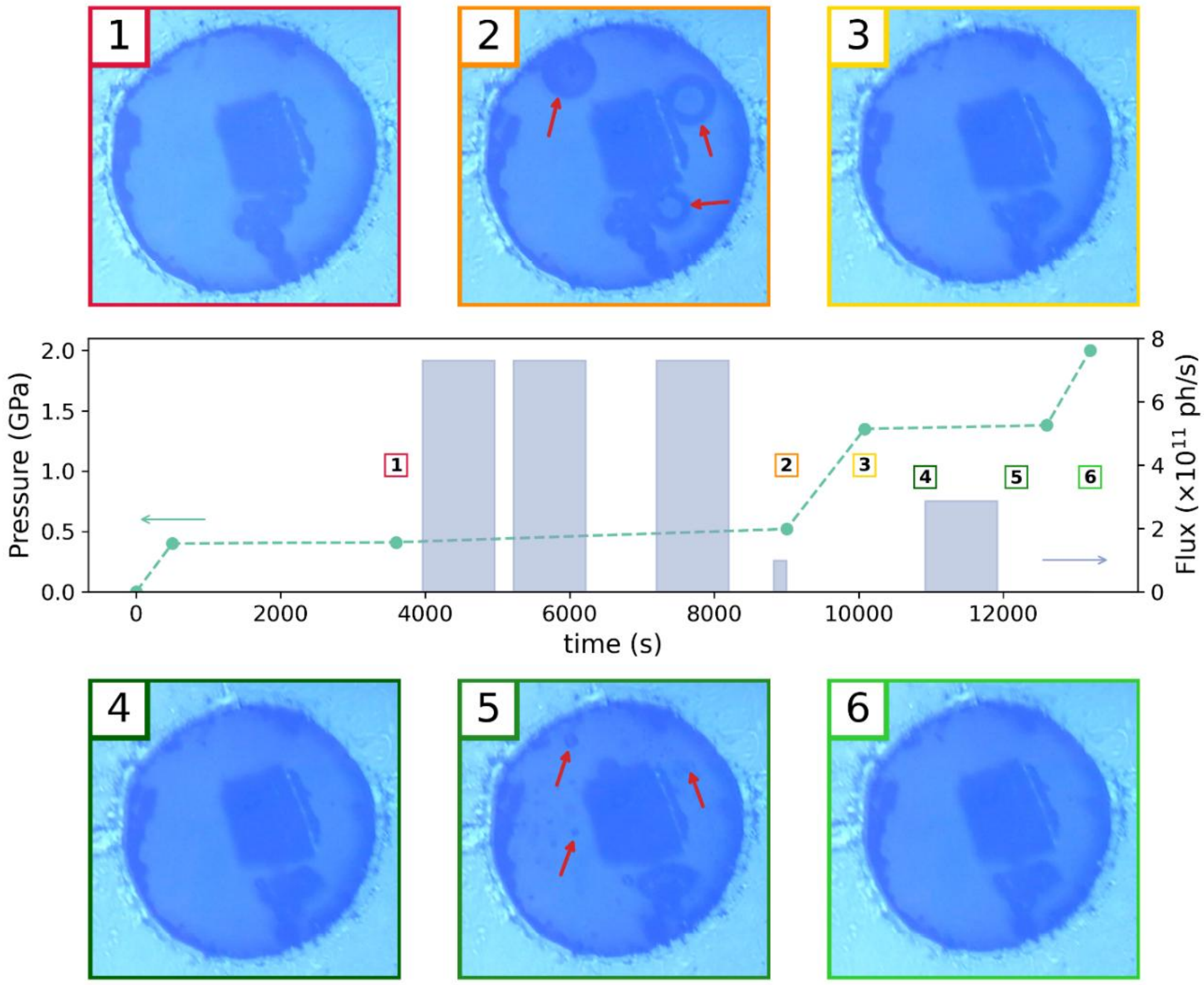

**Figure 6** Instantaneous flux on sample and evolution of the pressure measured on a ruby sphere, with micrographs of the experimental chamber in the gasket. The time of the acquisition of each micrograph is represented in the central panel by their corresponding labels. Red arrows highlight the position of bubbles.

The primary effect of this degassing is a lack of stability on the sample position. This in turn leads to artificial partial or full decorrelations and unphysical $g_2(q,\Delta t)$ functions. As the intrinsic dynamics of a glass can also be intermittent (Evenson *et al.*, 2015; Luo *et al.*, 2020), it is necessary to control the experimental volume before and after each scan to discard artificial sources for irregular dynamics within the glassy samples. Moreover, the apparition of bubbles in the 4:1 Methanol:Ethanol mixture implies broken bonds in the alcohol molecules due to the impinging x-rays. Not only this leads to the bubbles and the loss of sample stability mentioned above, but this also implies the creation of free radicals, which could later react with sample. At low pressure (< 2 GPa), it is therefore necessary to control the irradiation levels with an alcohol PTM to mitigate undesirable effects. Finally, the irradiation-induced degassing of the PTM stresses the importance of the online monitoring of the experimental volume, as other issues can appear during the experiment, such as a dendritic crystallization of 4:1 alcohol mixture at high pressure and temperature, and potential shear stress induced when the sample get pinched by the shrinking gasket.

### 3. Results: Stability, time resolution and physical aging under pressure

Once the sample stability has been achieved, it is possible to monitor the internal dynamics of a glass or a supercooled liquid at high pressure, as shown from our recent results (Cornet *et al.*, 2023) as well as from the TTCF in Fig. 5g.

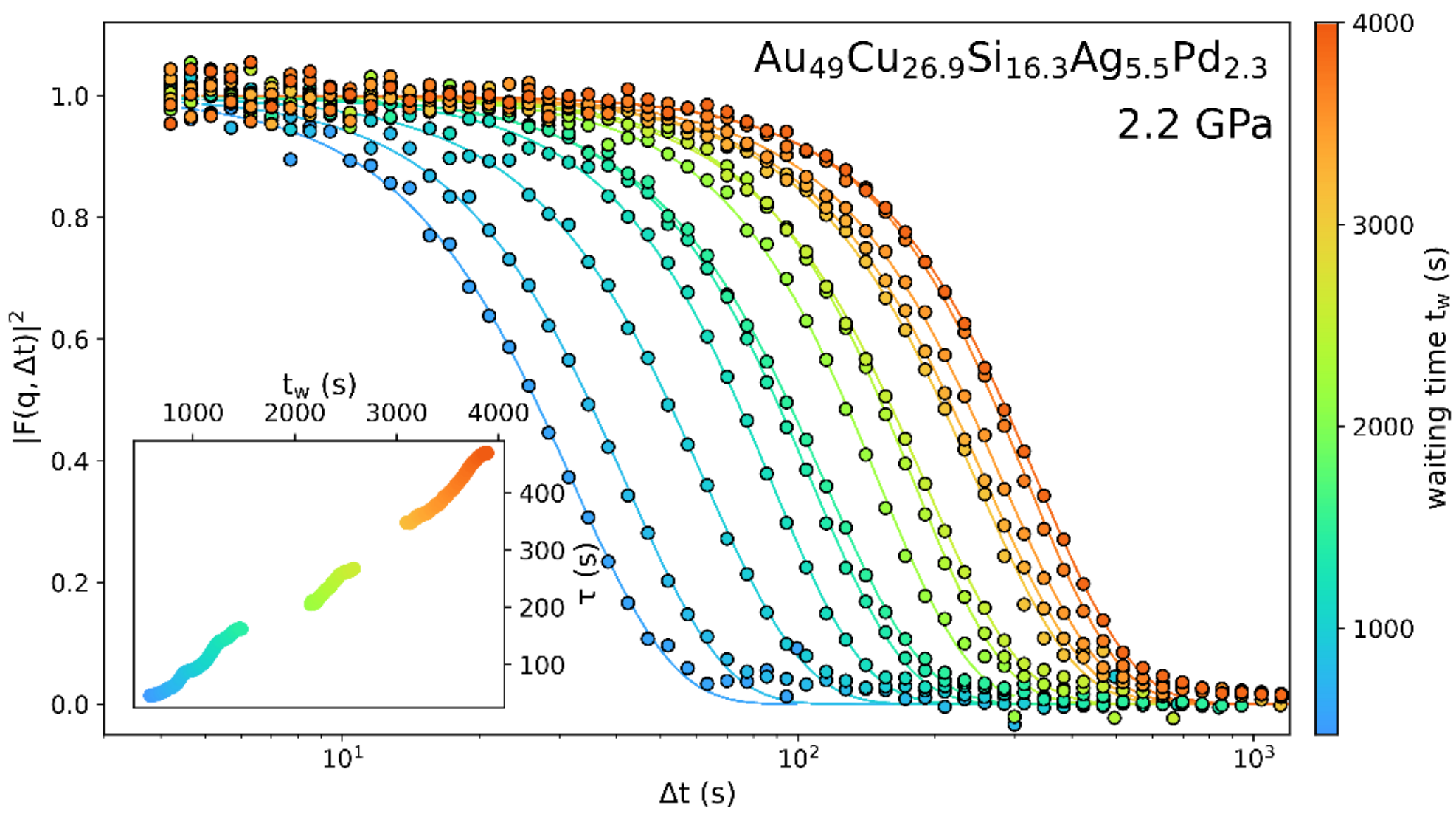

**Figure 7** Correlation functions in a $Au_{49}Cu_{26.9}Si_{16.3}Ag_{5.5}Pd_{2.3}$ metallic glass at 2.2 GPa, as a function of the elapsed time $t_w$ after setting the pressure. Lines are KWW fits to the data. The inset shows the evolution of the characteristic relaxation times τ from the KWW fits as a function of the waiting time $t_w$.

A sample stability compatible with XPCS measurements over long time scale has already been reported at high pressure and room temperature with a time resolution of 5s/frame (Zhang *et al.*, 2023), obtained before the EBS upgrade of the ESRF. The increased intensity of the coherent flux by two orders of magnitude allows measurement with integration time for the individual frames reduced by four orders of magnitude keeping the SNR constant, reaching the limit of 1 ms imposed by the frame rate of the detector. Thus, the achieved sample stability combined with the high SNR thanks to the enhanced coherent flux of the ESRF-EBS source opens the possibility for a time resolved evolution of the dynamic after the pressure perturbation and/or the temperature perturbation at high pressure.

The interest for time resolved XPCS under high pressure is highlighted on the Fig. 7, which features selected $|F(q,\Delta t)|^2$ functions obtained on a $Au_{49}Cu_{26.9}Si_{16.3}Ag_{5.5}Pd_{2.3}$ metallic glass at 2.2 GPa and 300K, as a function of

the waiting time $t_w$ after the pressure perturbation. These successive ISFs are obtained by binning the TTCF, effectively probing the correlation of stacks of 1000 frames of 0.1s exposure time with all the subsequent frames of the scan. The SNR of the ISFs remains excellent despite the relatively low number of frames, and a clear trend appears on the figure, where the shift of the ISF to larger times with increasing $t_w$ indicates a slowdown of the glass dynamics during the isobar. The characteristic relaxation time $\tau$ of the ISF can be extracted by fitting the Kohlrausch-Williams-Watts (KWW) function $|F(q,\Delta t)|^2 = e^{-2\times\left(\frac{\Delta t}{\tau(q)}\right)^{\beta(q)}}$ to the data. The final evolution of the characteristic time $\tau$ with $t_w$ is shown in the inset of the Fig. 7 with a time resolution of 100s (1000 frames × 0.1s), and a highly detailed curve with a point density of ten data points per second. Therefore, a real time resolved monitoring of the liquid and glassy dynamics is possible at high pressure while resolving the ISF from over 6 orders of magnitude of time, from $10^{-3}$ s to $10^{3}$ s.

## 4. Conclusion

A compression in the 0-10 GPa range translates to a density variation from 5 to 20% in metallic glasses, depending on the bulk modulus of the glass (Zeng *et al.*, 2014), to 34% in vitreous silica (Wakabayashi *et al.*, 2011), and to a similar variation in chalcogenide glasses (Mei *et al.*, 2006). Thus, the possibility to perform XPCS in-situ under high pressure and high temperature truly allows to investigate the effect of the density on the dynamical properties of the structural glasses and their corresponding supercooled liquids, including the glass transition. Some promising potentialities of HPHT-XPCS include the use of higher order correlation functions in order to probe the role of density, and therefore packing, on the dynamical heterogeneities in deeply supercooled liquids (Cipelletti *et al.*, 2002; Perakis *et al.*, 2017). The dynamical heterogeneities are variation in time and space of the dynamics in glass-formers, which appear during to the enormous increase of the viscosity upon cooling that eventually leads to the glass transition (Ludovic Berthier, Giulio Biroli, Jean-Philippe Bouchaud, Luca Cipelletti, Wim van Saarloos, 2011). XPCS can provide a quantitative estimate of this heterogeneity of the relaxation processes, through the determination of the $\chi_4$ four-probe correlation function (akin normalized variance) (Perakis *et al.*, 2017). As such, the development of HPHT-XPCS opens the possibility to monitor the intensity of the dynamical heterogeneities at the glass transition as a function of the liquid density. Another intriguing field that HP-XPCS can cover is polyamorphism, where amorphous systems switch between two distinct states (Tanaka, 2020). The advance of HPXPCS at $4^{th}$ generation synchrotron sources promises to enhance the foreseen possibilities offered by XPCS on polyamoprhism. HP-XPCS can also be applied beyond the field of liquids and glasses, as XPCS shows sensibility to the spin and/or charge fluctuations in quantum materials (Shpyrko *et al.*, 2007). As such, an adaptation of HP-XPCS to low temperatures has potential for the high temperature superconductors applications, where high pressure promotes the Curie temperature thanks to a redistribution of the charges (Jurkutat *et al.*, 2023).

**Acknowledgements** We acknowledge the ESRF (Grenoble, France) and DESY (Hamburg, Germany), a member of the Helmholtz Association HGF, for the provision of synchrotron radiation facilities. This research was carried out at the ID10 beamtime at ESRF under the LTP project HC4529 (DOI: 10.15151/ESRF-ES-517790385, 10.15151/ESRF-ES-816052788, 10.15151/ESRF-ES-1026320436 and 10.15151/ESRF-ES-1207223068), and beamtime was allocated at DESY for proposals I-20221106 EC, II-20230020 EC. We would also like to thank J. Jacobs, Y. Watier and K. Lhoste for assistance in using the ID10 beamline and high-pressure high-temperature support, and Daniel Weschke, Hanns-Pieter Liermann, Konstantin Glazyrin and Iris Schwark for assistance in using the P10 beamline and the high-pressure support. This project has received funding from the European Research Council (ERC) under the European Union's Horizon 2020 research and innovation programme (Grant Agreement No 948780). All data are available upon reasonable request.